\documentclass[12pt]{article}

\newcommand{\blind}{0}
\usepackage{amsmath, graphicx, enumerate, amsthm, bm}
\usepackage{natbib}  
\usepackage{algorithm}
\usepackage{algpseudocode}
\usepackage{setspace}
\usepackage{graphicx}
\usepackage{natbib}
\usepackage{booktabs} 
\usepackage{thmtools}

\usepackage{bookmark}
\usepackage{caption}
\usepackage{subcaption}
\usepackage{siunitx}
\usepackage{float} 
\usepackage{enumerate}
\usepackage{soul}
\usepackage{amsmath, amssymb} 
\usepackage{mathrsfs}
\usepackage{mhequ}
\usepackage{xcolor}
\usepackage{bbold} 
\usepackage{mathtools} 
\newtheorem{corollary}{Corollary}
\newtheorem{theorem}{Theorem}

\newtheorem{remark}{Remark}
  
\newtheorem{assumption}{Assumption} 
 
\newtheorem{definition}{Definition}
\usepackage{setspace} 
\usepackage{authblk}

\begin{document}

\if0\blind
{
  \title{\bf Learning Collapsed Patterns in Compositional Data: A Bayesian Heterogeneous Relative-Shift Approach}
  \author{Maoran Xu$^1$, Guanyu Hu$^2$\hspace{.2cm}\\
    \it\small $^1$Department of Statistics, Indiana University Bloomington. \\
    \it\small $^2$ Department of Statistics and Probability, Michigan State University.}
  \maketitle
} \fi

\if1\blind
{
  \bigskip
  \bigskip
  \bigskip
  \begin{center}
    {\LARGE\bf Learning Collapsed Patterns in Compositional Data: A Bayesian Heterogeneous Relative-Shift Approach}
\end{center}
  \medskip
} \fi

\bigskip
\begin{abstract}
Relative-shift regression provides a principled framework for modeling compositional covariates by quantifying how the response changes when mass is reallocated from one component to another. Yet many emerging compositional data problems extend beyond this classical setting, involving high-dimensional predictors and regression effects that vary across latent subpopulations. This complexity poses a dual challenge unmet by existing methods: recovering latent cluster structure while simultaneously achieving dimension reduction within each cluster. We propose a Bayesian heterogeneous relative-shift regression model that jointly learns latent clusters and parsimonious effect structures. Methodologically, we combine a projection-based shrinkage prior on identifiable contrasts, which induces exact coefficient ties within mixture components, with a mixture of finite mixtures prior that infers the number of clusters. Computationally, we develop a scalable hybrid MCMC algorithm that embeds a deterministic surrogate collapse operator within NUTS. Theoretically, we establish posterior consistency for both the latent partition and cluster-specific effect structures. Simulations confirm accurate recovery and strong predictive performance, and applications to cross-country macroeconomic data and spatial transcriptomics demonstrate the method's interpretability and practical utility.

\end{abstract}

\noindent%
\vfill

\newpage

\section{Introduction} \label{sec:intro}

Compositional predictor regression is central to empirical studies in which covariates represent parts of a whole, such as proportions or shares that sum to one. Such data arise naturally across economics \citep{fry2000compositional, franke2019different, meng2024bayesian}, public health \citep{leite2016applying, lubomski2022nutritional}, and genomics and microbiome research \citep{li2015microbiome, shi2016regression, zhang2021bayesian, jones2015adjusting, xia2013logistic, peterson2024analysis}. In economics, questions about sectoral contributions to GDP, trade shares, or household budget allocations are inherently compositional, and policy conclusions often depend on how mass shifts among parts \citep{meng2024bayesian}. In public health, dietary intake proportions, distributions of health expenditures, and compositional disease burdens are routinely analyzed to identify risk factors and guide interventions \citep{leite2016applying, lubomski2022nutritional}. In genomics, compositional structure appears in cell-type proportions and relative abundance profiles, especially in single-cell and microbiome studies \citep{jones2015adjusting, pflughoeft2012human, cao2019scdc}. Across these domains, the defining challenge is that inference must respect simplex geometry: increasing one component necessarily decreases at least one other component, and naive regression formulations can violate scale invariance or subcompositional coherence \citep{gloor2017microbiome}.

A major line of work addresses these constraints through log-ratio and log-contrast regression \citep{aitchison1984log, sun2020log, liu2022multivariate, han2023robust, zhao2024debiased}. Log-contrast models are attractive because they target subcompositional coherence and provide a principled regression framework for compositions. However, two limitations are especially consequential in modern applications. First, zeros and near-zeros are pervasive in high-throughput biological data, including microbiome and single-cell studies \citep{pflughoeft2012human, peterson2024analysis}. Log-ratio transforms require pseudo-counts, filtering, or other preprocessing choices that can introduce sensitivity and bias, and may complicate uncertainty quantification \citep{gloor2017microbiome}. Second, the log-ratio transformation maps compositions into an unconstrained space, but the resulting coefficients no longer correspond to interpretable single-component effects, and standard variable-selection or sparsity techniques developed for unconstrained regression are difficult to interpret on the simplex. These issues motivate models that bypass log-ratio transformations entirely and operate directly on proportions.

The relative-shift regression framework of \citet{li2023s} provides a robust approach to modeling compositional data by reparameterizing effects through pairwise coefficient contrasts, thereby ensuring full identifiability under the simplex constraint and avoiding the pseudo-count issues typical of log-ratio methods. While this framework successfully recovers approximately equi-sparse aggregation patterns via tree-guided regularization with post-hoc thresholding, its application has been limited to single, homogeneous populations. This homogeneity assumption is often too restrictive in practice; for instance, economic outcomes may reflect latent regional groupings with distinct sectoral drivers, and microbiome-disease dynamics may vary across unobserved clinical strata \citep{chang2023links, kononen2022prevotella, zhang2024bayesian}. Extending this framework to a mixture-of-regressions architecture can capture such heterogeneity, yet high-dimensional regimes where the predictor dimension matches or exceeds the sample size ($p \ge n$) introduce severe posterior instability. We formalize this challenge as \emph{high-dimensional posterior degeneracy}, noting that structurally constraining the component-specific coefficient space is essential for stability, whereas frequentist post-hoc thresholding remains incapable of delivering exact coefficient ties alongside calibrated uncertainty over the resulting aggregated structures.

Compositional effects frequently exhibit collapsed patterns, where multiple components share identical impacts on the response. For instance, distinct economic sectors may contribute similarly to aggregate output, or clusters of microbial taxa may share identical associations with a clinical outcome. While frequentist methods encourage such aggregation through fusion penalties \citep{tibshirani1996regression, xin2016efficient}, clustered regularization \citep{yuan2006model, she2010sparse}, supervised clustering of predictors \citep{bondell2008simultaneous, shen2010grouping, mehrotra2022simultaneous}, or tree-aggregated modeling \citep{bien2021tree, yan2021rare}, these approaches remain limited. They are generally designed for unconstrained predictors or single-population regressions, meaning they fail to address identifiability under simplex constraints, uncertainty quantification for aggregation decisions, or the joint learning of latent clusters alongside cluster-specific collapsed structures.

To address these limitations, a growing body of literature develops Bayesian methods for compositional regression, typically building on log-contrast models with shrinkage or selection priors \citep{zhang2021bayesian, zhang2024bayesian, peterson2024analysis}. However, these approaches focus strictly on a single homogeneous population and ignore heterogeneous composition-response relationships across latent subgroups. While Bayesian mixture modeling offers a natural framework to capture such heterogeneity and quantify uncertainty over both partitions and cluster-specific parameters \citep{escobar1995bayesian, neal2000markov, miller2018mixture}, it faces a fundamental pathology in high dimensions. As established by \citet{chandra2023escaping}, high-dimensional mixture posteriors over partitions tend to collapse onto either complete pooling or complete separation, irrespective of the clustering prior. This degeneracy is particularly severe in mixture-of-regressions settings with high-dimensional, unconstrained component coefficients that fail to stabilize the partition. Furthermore, traditional continuous shrinkage priors, such as the horseshoe \citep{carvalho2009handling} or Dirichlet-Laplace \citep{bhattacharya2015dirichlet}, shrink individual coefficients toward zero rather than enforcing the exact equality of contrasts that defines a collapsed compositional pattern. Even classical discrete spike-and-slab formulations \citep{george1993variable, ishwaran2005spike}, which can induce exact sparsity for isolated predictors, become computationally prohibitive in high dimensions and lack a natural mechanism to enforce structural ties among subsets of active coefficients. While Bayesian adaptations of the fused lasso \citep{kyung2010penalized} attempt to penalize pairwise differences, they generally produce continuous posteriors without exact coefficient fusion and struggle with identifiability under the simplex constraint. Applying these unconstrained priors within a mixture-of-regressions framework exacerbates these limitations, as they fail to provide the rigid structural regularization necessary to prevent the partition space from degenerating. Bridging Bayesian heterogeneity, high-dimensional posterior stability, and contrast-level aggregation under simplex constraints thus remains an unresolved methodological challenge.

Motivated by these gaps, we propose a Bayesian heterogeneous relative-shift regression framework for high-dimensional compositional predictors. Extending the relative-shift parameterization of \citet{li2023s}, our approach captures latent population heterogeneity through a finite mixture model while inducing exact coefficient ties within each component. Our contributions are threefold. Methodologically, we formalize a finite mixture-of-regressions framework for compositional data under general GLM likelihoods, where each mixture component employs a projection-based generalized $\ell_1$-ball prior on identifiable contrasts to induce equi-sparsity through exact coefficient ties, providing full posterior uncertainty over the latent partitions and cluster-specific aggregation structures. Theoretically, we adapt the high-dimensional posterior degeneracy framework of \citet{chandra2023escaping} to the compositional mixture setting, demonstrating that structural regularization within the contrast space mitigates posterior instability, and establish posterior consistency for the joint recovery of the number of clusters, the mixing measure, and the induced collapsed structures. Computationally, we develop a scalable MCMC algorithm that embeds a deterministic surrogate collapse operator within the No-U-Turn Sampler, sidestepping expensive iterative optimization during the projection step, and couple this sampler with search-based Bayesian clustering tools \citep{dahl2022search} to generate permutation-invariant partition summaries.

The remainder of this article is organized as follows. Section~\ref{sec:bhrs} develops the proposed Bayesian heterogeneous relative-shift regression framework, detailing the theoretical necessity of structural regularization, the generalized $\ell_1$-ball prior formulation, and the overarching mixture-of-finite-mixtures architecture. Section~\ref{sec:theory} provides a rigorous theoretical treatment of the model, establishing posterior consistency for both the mixing measure and the number of components; here, we verify that the compositional specification satisfies first- and second-order (strong) identifiability, ensuring optimal convergence rates within the MFM framework. Section~\ref{sec:bayes} outlines the scalable MCMC sampling strategy—including the deterministic surrogate collapse embedding for continuous trajectory simulation—alongside principled procedures for summarizing the posterior over partitions. Section~\ref{sec:sim} evaluates the operating characteristics and recovery performance of the proposed method through extensive synthetic experiments. Section~\ref{sec:app} presents our two motivating empirical applications, analyzing state-level sectoral economic drivers and spatial transcriptomic cell-type relationships, to demonstrate the model's capacity to discover interpretable, aggregated compositional structures in practice. Finally, Section~\ref{sec:discuss} concludes with a brief summary and a discussion of potential future extensions.

\section{Bayesian Heterogeneous Relative-Shift Regression}\label{sec:bhrs}

\subsection{Relative-shift regression for compositional predictors}
\label{sec:relative-shift}

Let $y = (y_1, \ldots, y_n)^\top$ denote a response vector for $n$ subjects.
For subject $i$, let $x_i = (x_{i1}, \ldots, x_{ip})^\top \in \mathcal{S}^{p-1}$
be a $p$-part composition residing on the simplex
$\mathcal{S}^{p-1} = \{x \in \mathbb{R}^p_+ : \sum_{j=1}^p x_j = 1\}$,
and let $c_i \in \mathbb{R}^q$ denote a vector of auxiliary (non-compositional)
covariates. We consider the generalized linear model
\begin{equation}
\label{eq:base-glm}
y_i \mid x_i, c_i \sim p(\cdot \mid \eta_i),
\qquad
\eta_i = c_i^\top \boldsymbol{\beta}_c + x_i^\top \theta,
\end{equation}
where $p(\cdot \mid \eta_i)$ denotes an exponential-family density with canonical
link, $\boldsymbol{\beta}_c \in \mathbb{R}^q$ collects the auxiliary-covariate
effects, and $\theta = (\theta_1, \ldots, \theta_p)^\top \in \mathbb{R}^p$
collects the compositional effects.

The simplex constraint $\sum_{j=1}^p x_{ij} = 1$ implies that the design
columns of the compositional block are linearly dependent: for any
constant $a \in \mathbb{R}$,
$x_i^\top \theta = x_i^\top (\theta + a \mathbf{1}_p)$, so $\theta$ is
identifiable only up to addition of a multiple of $\mathbf{1}_p$.
Consequently, individual coefficients $\theta_j$ carry no absolute meaning,
and shrinking a single coefficient toward zero does not correspond to a
well-defined notion of ``no effect'' for component $j$. What \emph{is}
identifiable is the contrast space spanned by pairwise differences
$\theta_\ell - \theta_j$ for $j \neq \ell$. Following \citet{li2023s},
we adopt the \emph{relative-shift} parameterization, in which inference
is conducted through these identifiable contrasts. For any pair of
features $j \neq \ell$, the joint contribution to the linear predictor
admits the reallocation decomposition
\begin{equation}
\label{eq:reallocation}
\theta_j x_{ij} + \theta_\ell x_{i\ell}
\;=\;
\theta_j (x_{ij} + x_{i\ell}) + (\theta_\ell - \theta_j)\, x_{i\ell},
\end{equation}
so the contrast $\theta_\ell - \theta_j$ measures the expected change in
the linear predictor under a unit reallocation of mass from component $j$
to component $\ell$, holding the remaining components fixed. The
relative-shift parameterization thus replaces the non-identifiable
question ``what is the effect of component $j$?'' with the identifiable
question ``what is the effect of shifting mass from $j$ to $\ell$?''.

Two implications of \eqref{eq:reallocation} guide the rest of our
development. First, when $\theta_\ell = \theta_j$, features $j$ and
$\ell$ exert indistinguishable effects on the response, and aggregating
them leaves the conditional likelihood invariant. Structural parsimony
on the compositional block therefore corresponds to \emph{equi-sparsity}:
exact equality of subsets of coefficients, rather than exact zeros at
individual coefficients. Second, the relative-shift formulation makes
no use of log-ratio transformations and operates directly on the raw
proportions $x_{ij}$, sidestepping the pseudo-count and zero-handling
issues that complicate inference under log-contrast models.

\subsection{Heterogeneous compositional effects and high-dimensional degeneracy}
\label{sec:heterogeneity}

The single-population model \eqref{eq:base-glm} assumes a single set of
compositional effects $\theta$ shared across all subjects. In many
applications this is too restrictive: economic outcomes may reflect
latent groupings of countries with distinct sectoral drivers, and
microbiome--disease associations may vary across unobserved subtypes
or clinical strata. To accommodate such heterogeneity, we extend
\eqref{eq:base-glm} to a mixture of relative-shift regressions. Let
$z_i \in \{1, \ldots, K\}$ denote a latent cluster assignment for
subject $i$, and let $\theta_k = (\theta_{k1}, \ldots, \theta_{kp})^\top$
be the compositional coefficient vector for cluster $k$. We assume the
auxiliary-covariate effects $\boldsymbol{\beta}_c$ are shared across
clusters, while the compositional effects vary:
\begin{equation}
\label{eq:mixture-glm}
y_i \mid x_i, c_i, z_i = k \sim p(\cdot \mid \eta_{i,k}),
\qquad
\eta_{i,k} = c_i^\top \boldsymbol{\beta}_c + x_i^\top \theta_k.
\end{equation}
This specification preserves the relative-shift interpretation within
each cluster: the contrast $\theta_{k\ell} - \theta_{kj}$ quantifies
the expected change in the linear predictor under a unit reallocation
of mass from feature $j$ to feature $\ell$ for subjects in cluster $k$.

A naive Bayesian implementation of \eqref{eq:mixture-glm} places
independent diffuse priors on each $\theta_k$. In modern compositional
applications, however, $p$ is often comparable to or larger than
$n$, and the per-cluster sample size $n/K$ can be small relative
to $p$. In this regime, unconstrained cluster-specific coefficients
lead to a well-documented pathology: the posterior over the latent
partition becomes unstable and tends to concentrate on extreme
configurations. The following result, adapted from
\citet{chandra2023escaping}, makes this phenomenon precise in the
mixture-of-regressions setting.

\begin{corollary}[High-dimensional partition degeneracy]
\label{cor:degeneracy}
Consider model \eqref{eq:mixture-glm} with independent priors on the
cluster-specific parameters $(\theta_k, \phi_k)$, and let $\Psi$ be
a partition of $\{(y_1 \mid x_1, c_1), \ldots, (y_n \mid x_n, c_n)\}$
with $\Psi'$ a coarsening of $\Psi$. If
\begin{equation}
\label{eq:ratio-pool}
\limsup_{p \to \infty}
\frac{\prod_{k \geq 1} \int \prod_{i : z_i = k} p(y_i \mid \eta_{i,k})\, d\Pi_0(\theta_k, \phi_k)}
     {\prod_{k \geq 1} \int \prod_{i : z'_i = k} p(y_i \mid \eta_{i,k})\, d\Pi_0(\theta_k, \phi_k)}
\;=\; 0,
\end{equation}
then the posterior concentrates on complete pooling,
$\Pi(z_1 = \cdots = z_n \mid Y) \to 1$. Conversely, if the limsup
in \eqref{eq:ratio-pool} equals $\infty$, the posterior concentrates
on complete separation, $\Pi(z_1 \neq \cdots \neq z_n \mid Y) \to 1$.
\end{corollary}

A proof of Corollary~\ref{cor:degeneracy} is
provided in Section~S1.1 of the Supplementary Materials. Corollary~\ref{cor:degeneracy} reveals that, absent structural
constraints on the cluster-specific coefficient space, the posterior
over partitions in high dimensions is essentially forced toward one
of two pathological extremes, with no stable intermediate regime in
which heterogeneous structures can be recovered. Crucially, the
result does not depend on the specific clustering prior (Dirichlet
process, finite mixture, mixture-of-finite-mixtures) or on the true
data-generating mechanism: the degeneracy is driven by the geometry
of the likelihood ratio in \eqref{eq:ratio-pool}, which in turn is
governed by the dimension and flexibility of $\theta_k$. Standard
shrinkage priors that shrink individual coefficients toward zero
(horseshoe, Dirichlet--Laplace, continuous spike-and-slab) reduce
the effective dimension of each $\theta_k$ but do not target the
relevant identifiable object: the contrast space spanned by
$\{\theta_{k\ell} - \theta_{kj}\}_{j \neq \ell}$. What is required
instead is a prior that regularizes the contrasts directly,
encouraging exact equality of subsets of coefficients within each
cluster and thereby reducing the effective dimension of the
compositional block in an identifiability-aware manner. We develop
such a prior in Section~\ref{sec:glbp}.

\subsection{The generalized $\ell_1$-ball prior}
\label{sec:glbp}
To regularize the contrasts of a coefficient vector directly, we adopt the generalized $\ell_1$-ball prior (GLBP) of \citet{xu2023bayesian}. Conceptually, the GLBP departs from standard continuous shrinkage priors, including the Bayesian lasso \citep{park2008bayesian}, horseshoe \citep{carvalho2009handling}, Dirichlet-Laplace \citep{bhattacharya2015dirichlet}, and continuous spike-and-slab \citep{rovckova2018spike}, which shrink individual coefficients toward zero but cannot enforce \emph{exact} equality among subsets of coefficients. The GLBP is constructed by combining a smooth precursor distribution with a deterministic projection onto a polyhedral constraint set, yielding a prior that places positive probability on faces of the constraint set and therefore on configurations in which many contrasts are exactly zero.

Let $\theta^* \in \mathbb{R}^p$ denote an unconstrained precursor with
prior
\begin{equation}
\label{eq:precursor}
\theta^* \sim \Pi_0 = \mathcal{N}(0, \tau_0^2 I_p).
\end{equation}
Given a contrast matrix $A \in \mathbb{R}^{m \times p}$ and a radius
$r > 0$, the GLBP-distributed parameter $\theta \in \mathbb{R}^p$ is
defined as the Euclidean projection of $\theta^*$ onto the polyhedral
ball
$\mathcal{B}_{A, r} = \{ z \in \mathbb{R}^p : \|A z\|_1 \leq r \}$,
\begin{equation}
\label{eq:projection}
\theta \;=\; P_{\mathcal{B}_{A, r}}(\theta^{*})
\;:=\; \mathop{\arg\min}_{z \,:\, \|A z\|_1 \leq r}\; \|\theta^* - z\|_2.
\end{equation}
We write $\theta \mid (A, r, \tau_0^2) \sim \mathrm{GLBP}(\Pi_0, A, r)$
for the induced prior on $\theta$. The matrix $A$ encodes the structural
relationships among coefficients along which equi-sparsity is desired.
For the relative-shift setting, $A$ is the
$\binom{p}{2} \times p$ matrix whose rows are the pairwise-difference
contrasts $e_\ell^\top - e_j^\top$ for $1 \leq j < \ell \leq p$, where
$e_j$ denotes the $j$th standard basis vector of $\mathbb{R}^p$, so that
$A z$ is the vector of all pairwise contrasts of $z$. The radius $r$
controls the overall complexity of the admissible $\theta$: smaller
$r$ forces more contrasts to be exactly zero, while larger $r$ allows
a richer set of distinct coefficient values.

Two properties of the projection \eqref{eq:projection} drive the
behavior of the prior. First, the constraint set
$\mathcal{B}_{A, r}$ is a polytope in $\mathbb{R}^p$ whose faces
correspond to subsets of the contrasts $\{A z\}_m$ being exactly zero;
projecting a continuous random variable onto a polytope places positive
probability on each face, so that $\Pr(\theta_\ell = \theta_j) > 0$
for every pair $(j, \ell)$ with $A$-row $e_\ell^\top - e_j^\top$.
Consequently, posterior draws of $\theta$ exhibit exact coefficient
ties with positive probability, in contrast to continuous shrinkage
priors which produce ties only in a measure-zero limit. Second, the
projection is a deterministic, almost-everywhere differentiable
function of $\theta^*$, so posterior inference on $\theta$ can be
carried out by sampling the smooth precursor $\theta^*$ and applying
\eqref{eq:projection} pointwise --- a computational advantage we exploit
in Section~\ref{sec:bayes}.

Mechanically, the GLBP can be viewed as a Bayesian counterpart to
fused-lasso-type regularization \citep{she2010sparse},
in which the penalty $\|A \theta\|_1$ on pairwise differences is
replaced by the hard constraint $\|A \theta\|_1 \leq r$. Unlike
penalized estimators, however, the GLBP yields a valid full posterior
distribution over $\theta$, propagating uncertainty about which
coefficients are tied. The radius $r$ is treated as a model
parameter: we place an exponential hyperprior $r \sim \mathrm{Exp}(a_r)$,
allowing the data to inform the overall degree of aggregation.

\subsection{Bayesian heterogeneous relative-shift regression}
\label{sec:full-model}

We now assemble the full model by placing a GLBP prior on the
cluster-specific compositional coefficients in the mixture-of-regressions
specification \eqref{eq:mixture-glm}, and equipping the latent partition
with a mixture-of-finite-mixtures (MFM) prior. 
For $i = 1, \ldots, n$ and $k \in \{1, \ldots, K\}$, the conditional
response model is
\begin{equation}
\label{eq:full-likelihood}
y_i \mid x_i, c_i, z_i = k \;\sim\; p(\cdot \mid \eta_{i,k}),
\qquad
\eta_{i,k} = c_i^\top \boldsymbol{\beta}_c + x_i^\top \theta_k,
\end{equation}
where $\boldsymbol{\beta}_c \in \mathbb{R}^q$ is shared across clusters
and $\theta_k \in \mathbb{R}^p$ is the cluster-specific compositional
effect vector.

\paragraph{Prior on compositional effects.}
Each cluster-specific coefficient vector $\theta_k$ is assigned an
independent GLBP prior with a shared contrast matrix $A$ encoding all
pairwise differences (Section~\ref{sec:glbp}) and a cluster-specific
radius $r_k$:
\begin{equation}
\label{eq:theta-prior}
\theta_k \mid r_k, A, \tau_0^2 \;\sim\; \mathrm{GLBP}(\Pi_0, A, r_k),
\qquad
r_k \;\sim\; \mathrm{Exp}(a_r),
\qquad
k = 1, \ldots, K,
\end{equation}
with precursor variance $\tau_0^2$. Allowing the radius to vary across
clusters lets the data determine the aggregation complexity of each
component independently, so that clusters with simple sectoral or
taxonomic structure can collapse to a small number of distinct
coefficient levels, while clusters with richer heterogeneity retain
finer resolution. The auxiliary-covariate coefficients receive a
diffuse Gaussian prior,
$\boldsymbol{\beta}_c \sim \mathcal{N}(0, \sigma_c^2 I_q)$ with
$\sigma_c^2 \sim \mathrm{IG}(\alpha_c, \beta_c)$.

\paragraph{Prior on the latent partition.}
The number of mixture components and the cluster assignments
$z = (z_1, \ldots, z_n)$ are governed by a mixture-of-finite-mixtures
(MFM) prior \citep{miller2018mixture}, which places
positive prior mass on each number of occupied components and yields a
consistent estimator of the true number of components under mild
regularity conditions. We adopt the exponential stick-breaking
realization of MFM: with rate $\lambda > 0$,
\begin{equation}
\label{eq:mfm}
\begin{aligned}
\varepsilon_k \mid \lambda &\stackrel{\mathrm{iid}}{\sim} \mathrm{Exp}(\lambda),
\qquad k = 1, 2, \ldots, \\
\tilde{K} &= \min\Bigl\{\, k \ge 1 : \textstyle\sum_{j=1}^{k} \varepsilon_j \ge 1 \,\Bigr\}, \\
\pi_k &= \varepsilon_k \;\; (k < \tilde{K}), \qquad
\pi_{\tilde{K}} = 1 - \textstyle\sum_{j < \tilde{K}} \varepsilon_j, \qquad
\pi_k = 0 \;\; (k > \tilde{K}), \\
z_i \mid \pi &\stackrel{\mathrm{iid}}{\sim} \mathrm{Categorical}(\pi_1, \pi_2, \ldots),
\qquad i = 1, \ldots, n.
\end{aligned}
\end{equation}
The stopping rule renders the number of active components $\tilde{K}$
an almost surely finite random variable, with the rate $\lambda$
governing its prior distribution. The MFM
specification is preferred over the more familiar Dirichlet process
mixture, which is known to overestimate the number of components
asymptotically \citep{miller2013simple}.

Together, equations \eqref{eq:full-likelihood}--\eqref{eq:mfm} define
the Bayesian heterogeneous relative-shift regression model. The
specification addresses the three challenges discussed in the
introduction in a unified manner: (i) the relative-shift parameterization
ensures that cluster-specific effects are interpreted through
identifiable mass reallocations, inheriting the compositional
identifiability of \citet{li2023s};
(ii) the GLBP prior \eqref{eq:theta-prior} regularizes the
contrasts of $\theta_k$ directly, producing exact coefficient
ties within each cluster and stabilizing posterior inference
against the high-dimensional partition degeneracy of
Corollary~\ref{cor:degeneracy}; and (iii) the MFM prior \eqref{eq:mfm} infers the number of latent
subpopulations $K$ from the data while propagating uncertainty
over the partition.  

\section{Theory}\label{sec:theory}

Posterior consistency for mixture-of-regressions models has been studied extensively by \citet{ho2022convergence} and \citet{do2025strong}. In this section, we establish posterior consistency for the proposed Bayesian heterogeneous relative-shift regression model under the Gaussian identity link $\mu_i = \mathbf{x}_i^\top\boldsymbol{\theta}_{z_i}$ with cluster-specific variance $\sigma_{z_i}^2$. We establish these results for the truncated representation used for posterior computation (Section~\ref{sec:bayes}), which caps the number of active components at $K_{\max} \ge K_0$; since the true number of clusters $K_0$ is finite, this over-fitted truncation captures the behavior of the infinite mixture prior of Section~\ref{sec:full-model} in the regime of interest.

We begin by introducing the notion of strong identifiability for mixture-of-regressions models and verifying that our compositional regression specification satisfies this condition. Identifiability of the kernel density ensures nonsingularity of the Fisher information matrix for the mixture distribution, which in turn guarantees that the mixing measure is learnable from the data. 

\begin{definition}
    Consider a general form of MoR model with
    \[f(y\mid x) = \sum_{j=1}^k p_j f(y\mid h_1(x,\boldsymbol{\theta}_{1j}), h_2(x,\boldsymbol{\theta}_{2j})),\]
    where $f$ is a dispersion exponential family distribution with parameter $\mu = h_1(x, \boldsymbol{\theta}_{1})$ and $\phi = h_2(x,\boldsymbol{\theta}_{2})$, and $\boldsymbol{\theta}_1\in\mathbb R^{d_1},\boldsymbol{\theta}_2\in\mathbb R^{d_2}$.  We say the family of conditional densities $f(y\mid h_1(x,\boldsymbol{\theta}_{1}), h_2(x,\boldsymbol{\theta}_{2}))$ is
    \begin{itemize}
        \item \textbf{first-order identifiable}, if for any $k\in\mathbb N$ and any given $k$ distinct elements $(\boldsymbol{\theta}_{11},\boldsymbol{\theta}_{21}),\ldots,(\boldsymbol{\theta}_{1k},\boldsymbol{\theta}_{2k})$,
\[\sum_{j=1}^k \alpha_j f_j (y\mid x) + \beta_j^\top \frac{\partial}{\partial\boldsymbol{\theta}_1}f_j(y\mid x) + \gamma_j^\top\frac{\partial}{\partial \boldsymbol{\theta}_2} f_j (y\mid x) =0 \]
holds for almost all $x, y$ only if $\alpha_j =0$, $\beta_j = \mathbf{0}^{d_1}$, and $\gamma_j = \mathbf{0}^{d_2}$.
        \item \textbf{second-order identifiable} (or \emph{strongly identifiable}), if for any given $k$ distinct elements $\left(\boldsymbol{\theta}_{11}, \boldsymbol{\theta}_{21}\right), \ldots,\left(\boldsymbol{\theta}_{1 k}, \boldsymbol{\theta}_{2 k}\right) \in \Theta_1 \times \Theta_2$ and any $s_1, \ldots, s_k \geq 1$, whenever the coefficients $\alpha_j \in \mathbb{R}, \beta_j \in \mathbb{R}^{d_1}, \gamma_j \in \mathbb{R}^{d_2}$, and $\rho_{j t} \in \mathbb{R}^{d_1}, \nu_{j t} \in \mathbb{R}^{d_2}$ for $j=1, \ldots, k$, $t=1, \ldots, s_j$ satisfy, for almost all $x, y$ (w.r.t.\ $\mathbb{P}_X \times \nu$),
$$
\begin{aligned}
\sum_{j=1}^k \alpha_j f_j(y \mid x) & +\beta_j^{\top} \frac{\partial}{\partial \boldsymbol{\theta}_1} f_j(y \mid x)+\gamma_j^{\top} \frac{\partial}{\partial \boldsymbol{\theta}_2} f_j(y \mid x)+\sum_{t=1}^{s_j}\left(\rho_{j t}^{\top} \frac{\partial^2}{\partial \boldsymbol{\theta}_1^2} f_j(y \mid x) \rho_{j t}\right) \\
& +\sum_{t=1}^{s_j}\left(\nu_{j t}^{\top} \frac{\partial^2}{\partial \boldsymbol{\theta}_2^2} f_j(y \mid x) \nu_{j t}\right)+\sum_{t=1}^{s_j}\left(\rho_{j t}^{\top} \frac{\partial^2}{\partial \boldsymbol{\theta}_1 \partial \boldsymbol{\theta}_2} f_j(y \mid x) \nu_{j t}\right)=0,
\end{aligned}
$$
it follows that $\alpha_j = 0$, $\beta_j = \mathbf{0}^{d_1}$, $\gamma_j = \mathbf{0}^{d_2}$, $\rho_{jt} = \mathbf{0}^{d_1}$, and $\nu_{jt} = \mathbf{0}^{d_2}$ for all $j, t$.
    \end{itemize}
    \label{def:strong_id}
\end{definition}

\begin{remark}\label{rem:id_our_model}
    The strong identifiability condition may or may not hold depending on the choice of link functions $h_1$, $h_2$ and distribution family $f$.
    \begin{itemize}
        \item The main model in our framework takes $f$ to be the normal distribution with $h_1(\mathbf{x},\boldsymbol{\theta}_j)=\mathbf{x}^\top\boldsymbol{\theta}_j$ and $h_2 =\sigma_j^2$. Because the mean is a non-trivial function of $\mathbf{x}$, the kernel density satisfies strong identifiability.
        \item If instead both the mean and variance in each cluster are constants, i.e., $h_1 = \boldsymbol{\theta}_j$, $h_2 = \sigma_j^2$, then the conditional densities satisfy only first-order identifiability and violate second-order identifiability, due to the heat equation $\frac{\partial f}{\partial\sigma^2} = \tfrac{1}{2}\frac{\partial^2 f}{\partial\mu^2}$ for the normal density.
        \item An intercept term in the regression model would introduce non-identifiability. 
Including a global intercept alongside compositional predictors that sum to one 
renders the model non-identifiable, as the intercept and compositional effects 
cannot be separately recovered. Compositional regression absorbs the intercept 
by nature: if $E(y_i\mid z_i) = \beta_0 + \mathbf{x}^\top\boldsymbol{\beta}$, 
then because $\sum_j x_{ij}=1$ we can equivalently write 
$E(y_i\mid z_i) = \mathbf{x}^\top(\boldsymbol{\beta} + \beta_0\mathbf{1})$, 
so the intercept is not separately identifiable and need not be included. 
Relying strictly on relative shifts avoids this pathology and ensures that 
cluster assignments and component-specific effects are uniquely determined.
    \end{itemize}
\end{remark}

Together, Definition~\ref{def:strong_id} and Remark~\ref{rem:id_our_model}
establish that the Gaussian relative-shift specification is strongly
identifiable, the structural property that underpins the consistency
results developed below.

\begin{assumption}\label{assump:regularity}
The following regularity conditions hold.
\begin{enumerate}
    \item \textbf{(Compact parameter space.)} The parameter space $\Theta = \Theta_\theta \times \Theta_\sigma$ is compact. 
    \item \textbf{(Bounded-below mixing weights.)} There exists a constant $\delta_\pi > 0$ such that the true mixing weights satisfy $\pi_{0k} \ge \delta_\pi$ for all $k = 1,\ldots,K_0$.
    \item \textbf{(Sub-Gaussian errors.)} The error distribution satisfies $\mathbb{E}[\exp(t\varepsilon_i)] \le \exp(\nu^2 t^2/2)$ for all $t\in\mathbb{R}$ and some $\nu > 0$.
    \item \textbf{(True complexity bounded by truncation.)} The true number of clusters satisfies $K_0 \le K_{\max}$, and the true mixing measure $G_0$ has all $K_0$ atoms distinct.
    \item \textbf{(Prior positivity near truth.)} The GLBP prior assigns positive probability to every $\ell_2$-neighborhood of $\boldsymbol{\theta}_{0k}$.
\end{enumerate}
\end{assumption}

\begin{theorem}\label{thm: consistency}
    Suppose the true distribution of the data $(y_i,\mathbf{x}_i)$ follows a regression model as specified and the data-generating process satisfies Assumption~\ref{assump:regularity}. Then, given any $G_0$ with the true number of clusters $K_0$, we have the following posterior consistency:
    \begin{itemize}
        \item Induced density convergence (in Hellinger distance $h$):
    \[\Pi^{(n)}\!\left[\,h(f_G, f_0) > C_1 \sqrt{\log n / n} \;\middle|\; \mathbf{y}^{(n)},\mathbf{x}^{(n)}\right]\to 0 \quad\text{a.s.\ under } \textstyle\prod_{i=1}^n \mathbb P_{G_0}.\]
    \item Number-of-clusters consistency: \[\Pi^{(n)}\!\left[K = K_0 \mid \mathbf{y}^{(n)},\mathbf{x}^{(n)}\right]\to 1\] a.s.\ under $\prod_{i=1}^n \mathbb P_{G_0}$.
    \item Parameter recovery in the Wasserstein-1 distance $W_1$: \[\Pi^{(n)}\!\left[\,W_1(G, G_0) > C_2 \sqrt{\log n / n} \;\middle|\; \mathbf{y}^{(n)},\mathbf{x}^{(n)}\right]\to 0\]
    in $\prod_{i=1}^n \mathbb P_{G_0}$-probability.
    \end{itemize}
\end{theorem}

A complete proof, together with the supporting lemmas it relies on, is provided in Section~S1.3 of the Supplementary Materials.

Theorem~\ref{thm: consistency} establishes that the proposed model is 
protected against a well-known pathology of Bayesian mixture models: the 
tendency to create spurious clusters that absorb noise as sample size grows. 
Under the structural regularization induced by the GLBP prior, the posterior 
probability of recovering the true number of clusters $K_0$ and the true 
coefficient structure converges to one as $n \to \infty$. This guarantee 
implies that identified clusters reflect genuine structural differences in 
the composition--response relationship rather than noise-induced fragmentation.
\section{Bayesian Inference}\label{sec:bayes}

\subsection{Bayesian computation}

Posterior sampling is implemented using the No-U-Turn Sampler (NUTS) via NumPyro \citep{hoffman2014no,phan2019composable}. While traditional implementations of the Mixture of Finite Mixtures (MFM) rely on custom Gibbs and surrogate Hamiltonian Monte Carlo (sHMC) schemes \citep{xu2023bayesian}, utilizing NUTS provides automatic step-size adaptation and trajectory-length tuning.

To maintain computational efficiency and differentiability within NUTS, inference utilizes a truncated representation with at most $K_{\max}$ active components, where $K_{\max}$ is a computational cap on the number of components rather than a modeling constraint---the prior in \eqref{eq:mfm} admits arbitrarily many components---and may, for example, be taken as the sample size or its logarithm. Because standard NUTS requires continuous parameter spaces, the discrete cluster assignments $z_i$ are analytically marginalized out during the log-density evaluation. The deterministic surrogate collapse operator (detailed in Section~\ref{sec:struct_mcmc}) is embedded directly within this evaluation step, allowing the HMC leapfrog integrator to traverse the structured posterior seamlessly (Algorithm~\ref{alg:nuts_eval}).

\begin{algorithm}[h]
\caption{Log-Posterior Evaluation for NUTS with GLBP Surrogate Collapse}
\label{alg:nuts_eval}
\begin{algorithmic}[1]

\Require Continuous state parameters: $\{\beta_k\}_{k=1}^{K_{\max}}$, $\{\sigma_k^2\}_{k=1}^{K_{\max}}$, and simplex weights $\boldsymbol{\pi}$.
\Ensure Log-posterior density $\log p(\Theta \mid \mathbf{y})$ and its gradients.
\Statex \textbf{Step 1: Structural Shrinkage (Deterministic Collapse)}
\For{$k = 1$ to $K_{\max}$}
    \State Sort $\beta_k$ to obtain order statistics $\beta_{k,(1)} \le \dots \le \beta_{k,(p)}$.
    \State Compute adjacent differences: $d_i = \beta_{k,(i+1)} - \beta_{k,(i)}$ for $i=1,\dots,p-1$.
    \State Apply soft-thresholding: $u_i = \max(d_i - \tau, 0)$.
    \State Reconstruct sorted sequence from the base $\tilde\beta_{k,(1)} = 0$: $\tilde\beta_{k,(i)} = \tilde\beta_{k,(i-1)} + u_{i-1}$ for $i = 2,\dots,p$.
    \State Apply inverse permutation to yield the structured coefficients $\tilde{\boldsymbol{\theta}}_k$.
    \State Re-center to the precursor level: $\tilde{\boldsymbol{\theta}}_k \gets \tilde{\boldsymbol{\theta}}_k - \overline{\tilde{\boldsymbol{\theta}}}_k + \overline{\boldsymbol{\beta}}_k$.
\EndFor
\Statex \textbf{Step 2: Marginalized Log-Likelihood Evaluation}
\State Initialize total log-likelihood $L = 0$.
\For{$i = 1$ to $n$}
    \State $\ell_i = \log \sum_{k=1}^{K_{\max}} \pi_k \, f(y_i \mid \mathbf{x}_i, \tilde{\boldsymbol{\theta}}_k, \sigma_k^2)$
    \State $L = L + \ell_i$
\EndFor
\Statex \textbf{Step 3: Gradient Computation}
\State Compute unnormalized log-posterior: $\log p(\Theta \mid \mathbf{y}) \propto L + \log p(\boldsymbol{\pi}) + \sum_{k} [\log p(\beta_k) + \log p(\sigma_k^2)]$.
\State Compute $\nabla \log p(\Theta \mid \mathbf{y})$ via automatic differentiation.
\State \Return $\log p(\Theta \mid \mathbf{y})$ and $\nabla \log p(\Theta \mid \mathbf{y})$
\end{algorithmic}
\end{algorithm}

\subsection{Efficient structural shrinkage}\label{sec:struct_mcmc}

The generalized $\ell_1$-ball prior (GLBP) projects coefficients onto an $\ell_1$-ball, functioning similarly to fused-lasso regularization for pairwise differences. As detailed in Algorithm~\ref{alg:nuts_eval}, we bypass expensive iterative optimization (e.g., ADMM) by applying one-sided soft-thresholding directly to the sorted adjacent differences of the coefficients:
\begin{equation}
u_i=\max(d_i-\tau,0).
\end{equation}
The threshold $\tau = \tau(r_k)$ is determined deterministically from the cluster-specific fusion radius $r_k$ as the smallest non-negative value such that the reconstructed sequence satisfies $\|A\tilde{\boldsymbol{\theta}}_k\|_1 \le r_k$. Reconstructing the sequence and applying the inverse permutation yields a blockwise-constant $\tilde{\boldsymbol{\theta}}_k$. This deterministic transformation allows NUTS to correctly target the structural shrinkage induced by the GLBP prior while scaling efficiently to high dimensions.

\subsection{Posterior inference for compositional effects}\label{sec:post_infer}

Because the compositional predictors $\mathbf{x}_i\in\mathbb{S}_{p-1}$ sum to 1 ($\sum_{j=1}^p x_{ij}=1$), coefficients are interpretable only through relative contrasts. Using posterior draws $\tilde{\boldsymbol{\theta}}^{(s)}$, we summarize effects as relative shifts \citep{li2023s}:
\begin{itemize}
    \item \textbf{Pairwise relative shifts:} $\Delta_{j\ell}=\tilde\theta_j-\tilde\theta_\ell$ represents the shift effect of reallocating one unit of mass from feature $\ell$ to $j$.
    \item \textbf{Group-level relative shifts:} For a collapsed group $\Omega$ of size $h$ with mean $\bar{\tilde\theta}_\Omega$, the contrast $\tilde\theta_j-\bar{\tilde\theta}_\Omega$ denotes shifting one unit of mass to component $j$, drawn evenly from all features in $\Omega$.
\end{itemize}
Directional non-negligibility is established when $(1-\alpha)100\%$ credible intervals exclude zero.

\subsection{Label-switching and partition summaries}\label{sec:label_switch}

Because finite mixture models are invariant to component label permutations, we resolve label-switching across the $S$ MCMC iterations using a three-step alignment procedure:
\begin{enumerate}
    \item \textbf{Posterior responsibilities:} Compute soft cluster assignments $\gamma_{ik}^{(s)} \propto \pi_k^{(s)} f(y_i \mid \mathbf{x}_i, \tilde{\boldsymbol{\theta}}_k^{(s)}, \sigma_k^{2(s)})$ for each iterate.
    \item \textbf{Dahl point estimate:} Identify a permutation-invariant partition by finding the iteration $s^\ast$ whose pairwise co-clustering matrix minimizes the squared distance to the posterior mean co-clustering matrix \citep{dahl2006model}.
    \item \textbf{Pivot-based relabeling:} Align all iterations to the pivot $s^\ast$ by minimizing $\sum_{k=1}^{K_{\max}} \sum_{i=1}^n (\gamma_{i,\sigma(k)}^{(s)} - \gamma_{ik}^{(s^\ast)})^2$ via the Hungarian algorithm \citep{kuhn1955hungarian}.
\end{enumerate}
Once aligned, valid posterior averages and credible intervals can be computed for all component-specific parameters. Finally, partition stability is validated using the Adjusted Rand Index \citep[ARI;][]{hubert1985comparing} and per-observation assignment entropy.

\section{Simulations}\label{sec:sim}

\subsection{Simulation Setup}

We evaluate the proposed heterogeneous relative-shift regression model under the Gaussian identity link, testing its ability to recover latent clusters, estimate collapsed coefficient structures, and maintain stable posterior inference under varying signal strengths and sample sizes. To extend the realism of these simulations, the blockwise constant patterns are designed to mirror the correlated, low-dimensional groupings often found in real compositional data (e.g., co-expressed biological pathways or interlinked economic sectors).

\textbf{Covariates.} For each observation $i$, the compositional covariate $\mathbf{x}_i$ is generated from a symmetric Dirichlet distribution:
\( 
\mathbf{x}_i \sim \mathrm{Dirichlet}(\mathbf{1}_p).
\)
The dimension $p$ varies by experiment (see Section~\ref{sec:sim_clean} and Section~\ref{sec:sim_hardcluster}). The same marginal distribution is shared across all clusters, ensuring the latent partition is driven entirely by coefficient heterogeneity rather than covariate clustering. \textbf{Latent Clusters \& Likelihood.} We simulate $K_0=3$ equal-sized latent clusters ($n_k = n/3$) with deterministic block assignments $z_i \in \{1,2,3\}$. The response is generated via $y_i = \mathbf{x}_i^\top\boldsymbol{\theta}_{z_i}+\varepsilon_i$, where $\varepsilon_i\sim \mathcal{N}(0,\sigma^2)$. \textbf{Collapsed Coefficient Structure.} Each cluster-specific coefficient vector $\boldsymbol{\theta}_k$ is a piecewise-constant profile, so that distinct components of the composition share identical effects within blocks. The exact per-block values and number of plateaus are specified in the experiment-specific subsections below; they are designed so that (i) the Dirichlet-averaged cluster means $\mathbb{E}[\mathbf{x}_i^\top \boldsymbol{\theta}_k]$ are well separated and (ii) the equi-sparse plateau structure matches the regime targeted by the GLBP prior. \textbf{Implementation.} Models are fit using the No-U-Turn Sampler \citep[NUTS;][]{hoffman2014no} in NumPyro \citep{phan2019composable} (2000 warmup, 3000 posterior samples). The MFM prior uses an exponential stick-breaking construction; the GLBP prior uses an $r_k \sim \mathrm{Exp}(a_r)$ hyperprior on the collapse radius. \textbf{Post-processing.} To address label-switching, we use the Dahl least-squares summary \citep{dahl2006model} on posterior responsibilities to find a permutation-invariant point estimate. All MCMC samples are then aligned to a common reference using the Hungarian algorithm for valid parameter averaging. The following evaluation metrics are averaged over $R$ replicates:
\begin{itemize}
    \item \textbf{ARI}: Adjusted Rand Index between the Dahl point estimate and the true partition ($1$ = perfect, $0$ = chance).
    \item \textbf{CoeffMSE}: Mean squared error of the relabeled posterior mean coefficients, $\mathrm{CoeffMSE} = K_0^{-1}\sum_{k=1}^{K_0}\|\hat{\boldsymbol{\theta}}_k - \boldsymbol{\theta}_{0k}\|_2^2$.
    \item \textbf{$K$ recovery}: Fraction of replicates estimating exactly the true number of clusters.
\end{itemize}

\subsection{Signal Recovery Under Exact Cluster Recovery}\label{sec:sim_clean}

First, we isolate coefficient recovery by using a ``clean'' regime where both models recover the latent partition near-perfectly. 

We set $K_0=3$ clusters, $n=600$, and $p=20$. The coefficients at scale $a > 0$ are piecewise-constant profiles with distinct Dirichlet-averaged means:
\begin{align*}
\boldsymbol{\theta}_1 &= (a\,\mathbf{1}_{p/2},\, 3a\,\mathbf{1}_{p/2}), \quad & \mathbb{E}[\mathbf{x}_i^\top\boldsymbol{\theta}_1] &= 2a. \\
\boldsymbol{\theta}_2 &= (-a\,\mathbf{1}_{p/2},\, -3a\,\mathbf{1}_{p/2}), \quad & \mathbb{E}[\mathbf{x}_i^\top\boldsymbol{\theta}_2] &= -2a. \\
\boldsymbol{\theta}_3 &= (a\,\mathbf{1}_{p/4},\, -a\,\mathbf{1}_{p/2},\, a\,\mathbf{1}_{p/4}), \quad & \mathbb{E}[\mathbf{x}_i^\top\boldsymbol{\theta}_3] &= 0.
\end{align*}

We evaluate two signal strengths (a scale factor of $2.0$ with noise $0.3$, and a scale factor of $2.5$ with noise $0.5$). Tuning parameters are fixed at $\lambda=4.0$, $K_{\max}=4$, and $a_r=0.5$ across $R=50$ replicates.

\begin{table}[ht]
\centering
\small
\caption{Clean-regime simulation ($K_0=3$, $n=600$, $p=20$, $R=50$).}
\label{tab:sim_clean}
\begin{tabular}{llcccc}
\toprule
Setting & Model & Exact Recovery & Mean ARI & CoeffMSE Mean $[95\% \mathrm{CI}]$ & PredMSE \\
\midrule
$a=2.0, \sigma=0.3$ & GLBP & $50/50$ & $1.0000$ & $0.147\ [0.038, 0.374]$ & $0.095$ \\
$a=2.5, \sigma=0.5$ & GLBP & $47/50$ & $0.9997$ & $0.428\ [0.109, 0.974]$ & $0.255$ \\
$a=2.5, \sigma=0.5$ & Baseline & $46/50$ & $0.9996$ & $1.648\ [1.133, 2.193]$ & $0.273$ \\
\bottomrule
\end{tabular}
\end{table}

As shown in Table~\ref{tab:sim_clean}, both methods easily resolve the overarching cluster structure. However, at the higher noise level, the GLBP model reduces the mean coefficient MSE by a factor of $3.85$ compared to the unconstrained Baseline, yielding strictly non-overlapping $95\%$ confidence intervals. By pooling information across covariates within the same fused block, GLBP effectively immunizes the parameter estimates against observation noise that causes the Baseline model to overfit. 

To verify this statistical efficiency scales advantageously into higher dimensions, we repeated the $a=2.5, \sigma=0.5$ configuration at $p \in \{30, 50\}$. 

\begin{table}[ht]
\centering
\small
\caption{Clean-regime scaling ($a=2.5, \sigma=0.5, n=600, R=50$). }
\label{tab:sim_clean_scaling}
\begin{tabular}{cccc}
\toprule
$p$ & GLBP CoeffMSE $[95\% \mathrm{CI}]$ &  Baseline CoeffMSE $[95\% \mathrm{CI}]$ & Ratio   \\
\midrule
$20$ & $0.428\ [0.109, 0.974]$ & $1.648\ [1.133, 2.193]$ & $3.85\times$   \\
$30$ & $2.221\ [0.964, 3.712]$ & $3.955\ [2.817, 5.349]$ & $1.78\times$   \\
$50$ & $6.102\ [5.477, 6.985]$ & $10.901\ [8.899, 13.812]$ & $1.79\times$   \\
\bottomrule
\end{tabular}
\end{table}

Table~\ref{tab:sim_clean_scaling} confirms that while both models consistently recover the exact clusters, GLBP structurally limits degrees of freedom. In paired replicate comparisons, GLBP achieved strictly lower MSE in nearly every instance, confirming that structural regularization consistently limits estimation 
error as dimensionality grows.

\subsection{Cluster Recovery Under Sample Scarcity}\label{sec:sim_hardcluster}

This regime mimics settings common in biological assays and macroeconomic 
analyses, where the per-cluster sample size $n/K_0$ approaches the ambient 
dimension $p$.

\begin{table}[ht]
\centering
\small
\caption{Cluster-scarcity regime ($p=30, n=120, a=1.5, \sigma=0.5, R=50$).}
\label{tab:sim_hardcluster}
\begin{tabular}{lcccccc}
\toprule
Model & $K=K_0$ & Mean ARI & ARI $95\%$ CI & Mean CoeffMSE & CoeffMSE $95\%$ CI & Mean PredMSE \\
\midrule
GLBP     & $42/50$ & $0.980$ & $[0.925, 1.000]$ & $2.46$  & $[2.20, 3.06]$   & $0.327$ \\
Baseline & $1/50$  & $0.572$ & $[0.409, 0.744]$ & $54.25$ & $[34.84, 75.61]$ & $0.342$ \\
\bottomrule
\end{tabular}
\end{table}

Table~\ref{tab:sim_hardcluster} demonstrates that GLBP dramatically improves both clustering accuracy and signal recovery in highly sparse regimes. The unregularized Baseline model fails to recover the true partition, 
typically assigning data to four or more clusters because high-variance 
coefficient estimates cause random noise to be misidentified as structural 
separation between cluster means.

By shrinking the effective parameter dimension down to just the true number of piecewise levels, the GLBP prior prevents this noise-induced fragmentation. It concentrates posterior mass on the correct $K_0 = 3$ clusters in $84\%$ 
of replicates and achieves a $22$-fold reduction in coefficient MSE relative 
to the Baseline.

\section{Real Examples}\label{sec:app}

\subsection{World GDP per Capita and Sectoral Composition}

We illustrate the proposed model using 2019 data on 105 countries to examine how sectoral composition is associated with log PPP-adjusted GDP per capita. The main objective is to assess whether national output composition provides additional explanatory information beyond broad macroeconomic and demographic differences. To separate these two sources of variation, we include two groups of covariates.

The first group contains seven standardized non-compositional variables: life expectancy, population growth, primary school enrollment, trade openness, urban population share, internet use, and unemployment rate. These variables capture general differences in development, population dynamics, openness, and labor-market conditions. They enter the model through a shared global coefficient vector $\boldsymbol{\beta}_c$, so their effects are assumed to be common across all countries.

The second group consists of 18 ISIC sectoral output shares, such as Manufacturing, Construction, Education, and other production sectors. Since these shares sum to one, they are compositional predictors. They enter through cluster-specific coefficient vectors $\boldsymbol{\theta}_k$, allowing the relationship between sectoral structure and GDP per capita to vary across latent groups of countries. This is important because the same sectoral share may have different implications for income depending on the broader economic structure of the country.

We fit the Gaussian identity-link instance of the model in Section~\ref{sec:full-model}, taking the response $y_i$ to be the log PPP-adjusted GDP per capita of country $i$:
\[
y_i \mid z_i = k \;\sim\; \mathcal{N}\!\left(\mathbf{c}_i^\top \boldsymbol{\beta}_c + \mathbf{x}_i^\top \boldsymbol{\theta}_k,\; \sigma_k^2\right),
\]
where $\mathbf{c}_i$ collects the seven non-compositional covariates with shared effect $\boldsymbol{\beta}_c$, $\mathbf{x}_i$ the 18 sectoral shares with cluster-specific effect $\boldsymbol{\theta}_k$, and $\sigma_k^2$ is a cluster-specific error variance.

To evaluate the additional contribution of the compositional predictors, we compute
\[
R^2_{\text{comp,increment}} = R^2_{\text{full}} - R^2_{\text{noncomp}}.
\]
This measures the extra variation in log GDP per capita explained by sectoral composition after the non-compositional variables have already been included.

\begin{table}[ht]
\centering
\small
\caption{Posterior-averaged $R^2$ decomposition and clustering summaries. Mean VI is the average Variation of Information distance from posterior draws to the Dahl partition.}
\label{tab:r2_compare_combined}
\begin{tabular}{lcccc}
\toprule
Model & $R^2_{\text{full}}$ & $R^2_{\text{comp,increment}}$ & $K$ (active clusters) & Mean VI \\
\midrule
Baseline (plain mixture) & $0.920$ & $0.135$ & $4$ & $2.04$ \\
Proposed (GLBP prior) & $0.914$ & $0.064$ & $4$ & $1.61$ \\
\bottomrule
\end{tabular}
\end{table}

\begin{figure}[htbp]
    \centering
    \includegraphics[width=0.85\textwidth]{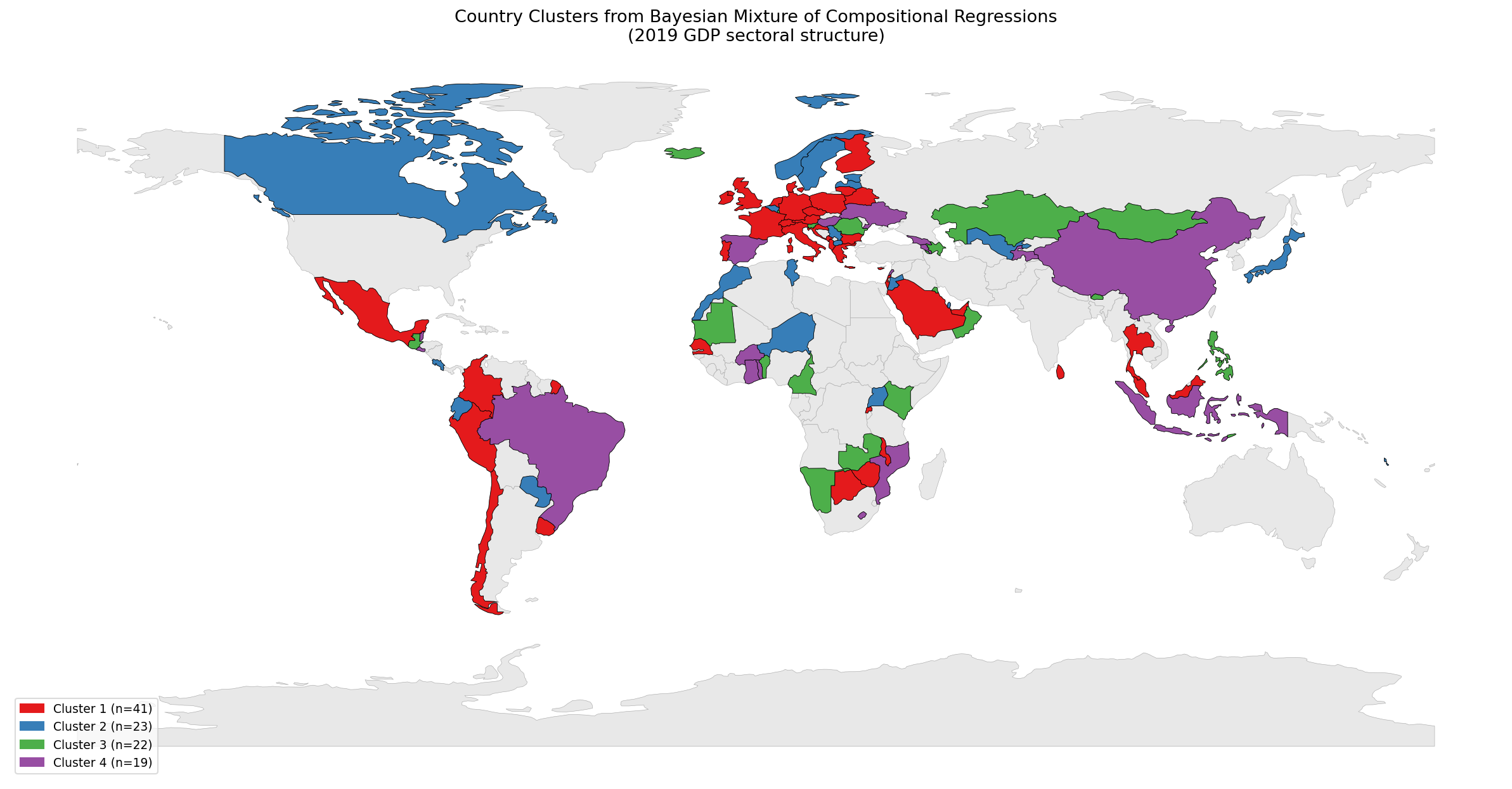}
    \caption{Estimated cluster assignments for the 105 countries under the proposed model, based on the Dahl partition.}
    \label{fig:gdp_worldmap}
\end{figure}

\begin{figure}[htbp]
    \centering
    \includegraphics[width=0.85\textwidth]{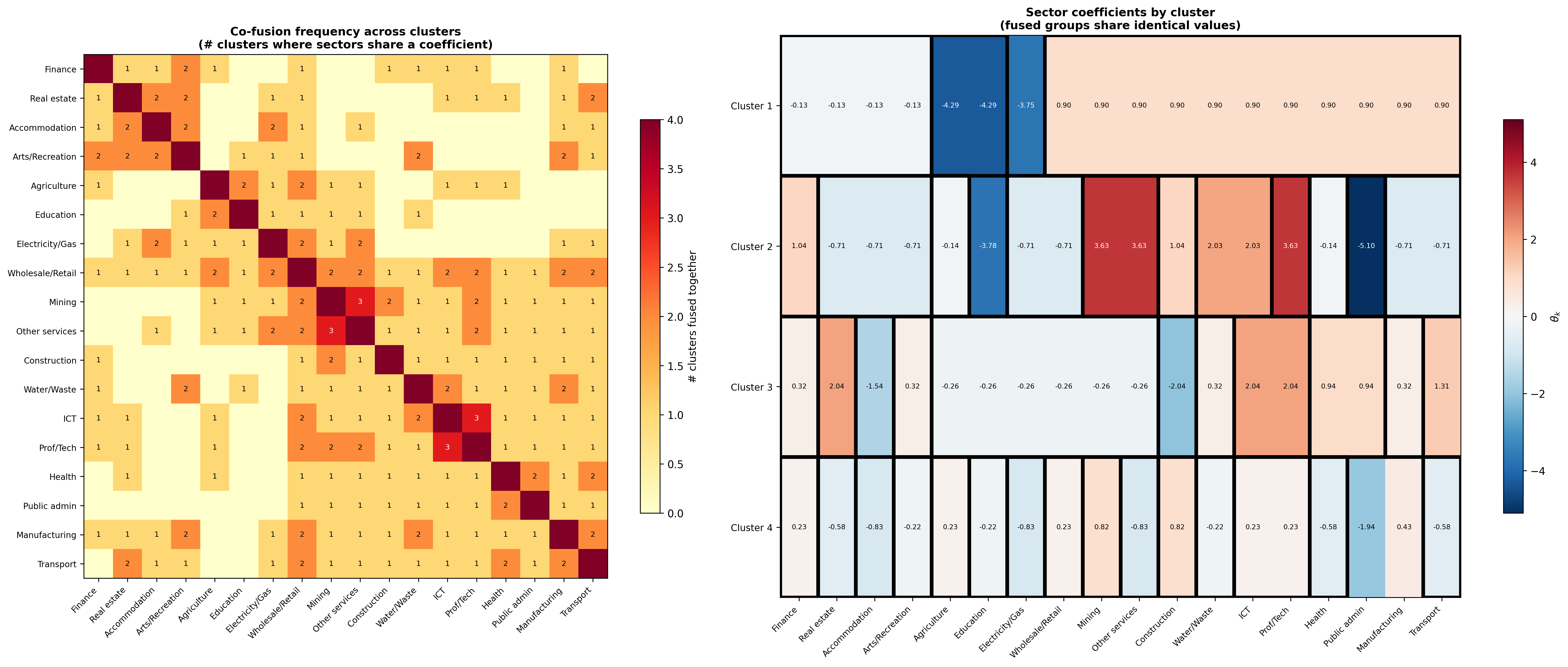}
    \caption{Left: posterior co-fusion frequencies for sector pairs across clusters. Right: cluster-specific posterior mean coefficient values.}
    \label{fig:merge_pattern}
\end{figure}

Table~\ref{tab:r2_compare_combined} shows that the proposed GLBP model achieves almost the same overall fit as the baseline mixture model. The full $R^2$ decreases only slightly from $0.920$ to $0.914$, indicating that the additional regularization imposed by the GLBP prior does not meaningfully reduce predictive performance. The main difference appears in the incremental $R^2$ attributed to the compositional covariates. The baseline model assigns an additional $0.135$ of explanatory power to the sectoral shares, while the proposed model assigns $0.064$. This smaller value does not imply that sectoral composition is irrelevant. Instead, it suggests that the GLBP prior removes redundant variation among the 18 sectoral shares and retains only the more stable low-dimensional sectoral signal. In this sense, the proposed model gives a more conservative and interpretable estimate of the contribution of economic structure.
Both models identify four active clusters, but the proposed model produces more stable clustering. The mean VI decreases from $2.04$ under the baseline to $1.61$ under the GLBP prior. Since lower VI indicates stronger agreement between posterior partitions and the representative Dahl partition, this result suggests that the proposed prior reduces posterior uncertainty in the clustering structure.
Figure~\ref{fig:gdp_worldmap} displays the estimated country-level partition. The clusters are spread across different continents rather than being concentrated within single geographic regions. This indicates that the model is not simply grouping countries by location. Instead, it identifies countries with similar relationships between sectoral composition and GDP per capita, even when those countries are geographically distant.
The GLBP prior also provides a more interpretable representation of the sectoral effects. Rather than estimating 18 unrelated sector-specific coefficients within each cluster, the prior encourages coefficients with similar effects to fuse together. As a result, the estimated $\hat{\boldsymbol{\theta}}_k$ values form horizontal plateaus, where several sectors share the same or nearly the same association with GDP per capita. Figure~\ref{fig:merge_pattern} illustrates this shrinkage pattern. The left panel shows posterior co-fusion frequencies for sector pairs, indicating which sectors are most frequently merged across posterior draws. The right panel shows the cluster-specific posterior mean coefficient values. Together, these plots reveal not only that fusion occurs, but also which sectors are grouped together and how the fused sectoral effects differ across clusters.

\begin{table}[ht]
\centering
\small
\caption{Within-cluster summaries for the proposed model. Effective dimension is the number of distinct fused coefficient levels in the posterior mean $\hat{\boldsymbol{\theta}}_k$.}
\label{tab:cluster_metrics}
\begin{tabular}{lccccc}
\toprule
Cluster & $N$ & Mean GDP (PPP) & RMSE & MAE & Eff.\ Dim \\
\midrule
1 & 41 & 20{,}232 & 0.333 & 0.261 & 9 \\
2 & 23 & 20{,}562 & 0.463 & 0.276 & 5 \\
3 & 22 & 18{,}465 & 0.521 & 0.395 & 7 \\
4 & 19 & 12{,}129 & 0.363 & 0.297 & 5 \\
\bottomrule
\end{tabular}
\end{table}

Table~\ref{tab:cluster_metrics} gives additional detail on the four clusters. Cluster sizes range from 19 to 41 countries, suggesting that the partition is reasonably balanced. The clusters differ in average GDP, but they are not simply ordered by income. For example, Clusters 1 and 2 have similar mean GDP values, yet they differ in prediction error and effective dimension. This suggests that the latent groups capture differences in the sectoral-income relationship, not only differences in income level.
The prediction errors also vary across clusters. Cluster 1 has the smallest RMSE and MAE, indicating that the model fits this group particularly well. Cluster 3 has the largest errors, suggesting greater unexplained heterogeneity among its countries. This may reflect factors not included in the model, such as institutional quality, natural resource dependence, or differences in human capital beyond primary school enrollment.
The effective dimension column highlights the parsimony gained from the GLBP prior. Although each cluster contains 18 original sectoral coefficients, the fitted model reduces them to only 5 to 9 distinct coefficient levels. Clusters 2 and 4 are especially compact, with only 5 fused levels each, while Cluster 1 retains a richer structure with 9 levels. Thus, the model adapts the amount of sectoral complexity to each cluster rather than imposing the same level of shrinkage everywhere.

Overall, the proposed GLBP model provides a strong balance between fit, stability, and interpretability. It maintains nearly the same predictive accuracy as the baseline model, improves posterior clustering stability, and substantially reduces the dimension of the sectoral coefficient structure. The results suggest that sectoral composition does contain meaningful information about GDP per capita, but this information can be summarized through a smaller number of fused sectoral effect levels rather than 18 fully distinct effects in every cluster.

\subsection{Spatial Omics Data: Cell Type Composition and Gene Expression}
\label{sec:transcriptomic}

We next apply the model to a spatial transcriptomics dataset from a human lung adenocarcinoma tissue section generated using the 10x Genomics Visium platform. The dataset contains expression measurements for 18{,}066 genes across $n = 3{,}813$ spatial spots. Using CARD deconvolution \citep{ma2022spatially}, each spot is represented by a 24-dimensional cell type proportion vector, covering immune, tumor, stromal, and epithelial lineages. This setting is well suited to the proposed model because the predictors are compositional and because the relationship between cell type composition and gene expression may vary across different regions of the tumor microenvironment.

We focus on the gene \emph{SPP1}, also known as osteopontin, which is widely associated with tumor microenvironment remodeling, immune regulation, and tumor progression. Let $y_i$ denote the observed UMI count for \emph{SPP1} at spot $i$, and let $\mathbf{x}_i$ denote the 24-dimensional cell type proportion vector. To account for differences in sequencing depth across spots, we include an offset $t_i$ equal to the total UMI count at spot $i$:
\[
y_i \mid z_i = k \sim \mathrm{Poisson}(\lambda_{ik}), \qquad
\log \lambda_{ik} = \log t_i + \alpha_k + \mathbf{x}_i^\top \boldsymbol{\theta}_k .
\]
Here, $\alpha_k$ is a cluster-specific intercept, while $\boldsymbol{\theta}_k$ captures how cell type composition is associated with normalized \emph{SPP1} expression within cluster $k$. The GLBP prior is placed on pairwise-difference contrasts of $\boldsymbol{\theta}_k$, encouraging cell types with similar expression effects to share the same coefficient value. Thus, the model simultaneously identifies latent spatial-expression regimes and simplifies the cell type effect structure within each regime.
The model achieves a high posterior-averaged pseudo-$R^2$ of $0.904$, indicating that the combination of cell type composition, sequencing-depth adjustment, and latent clustering explains a large proportion of the variation in \emph{SPP1} counts. The posterior concentrates on $K=6$ active clusters. Two large clusters, with sizes 1{,}679 and 1{,}482, cover most of the tissue section, suggesting that much of the sample follows two dominant composition--expression regimes. In contrast, the four smaller clusters, with sizes ranging from 48 to 229 spots, capture more localized patterns that would likely be missed by a single global regression model.

\begin{figure}[htbp]
    \centering
    \includegraphics[width=.35\textwidth]{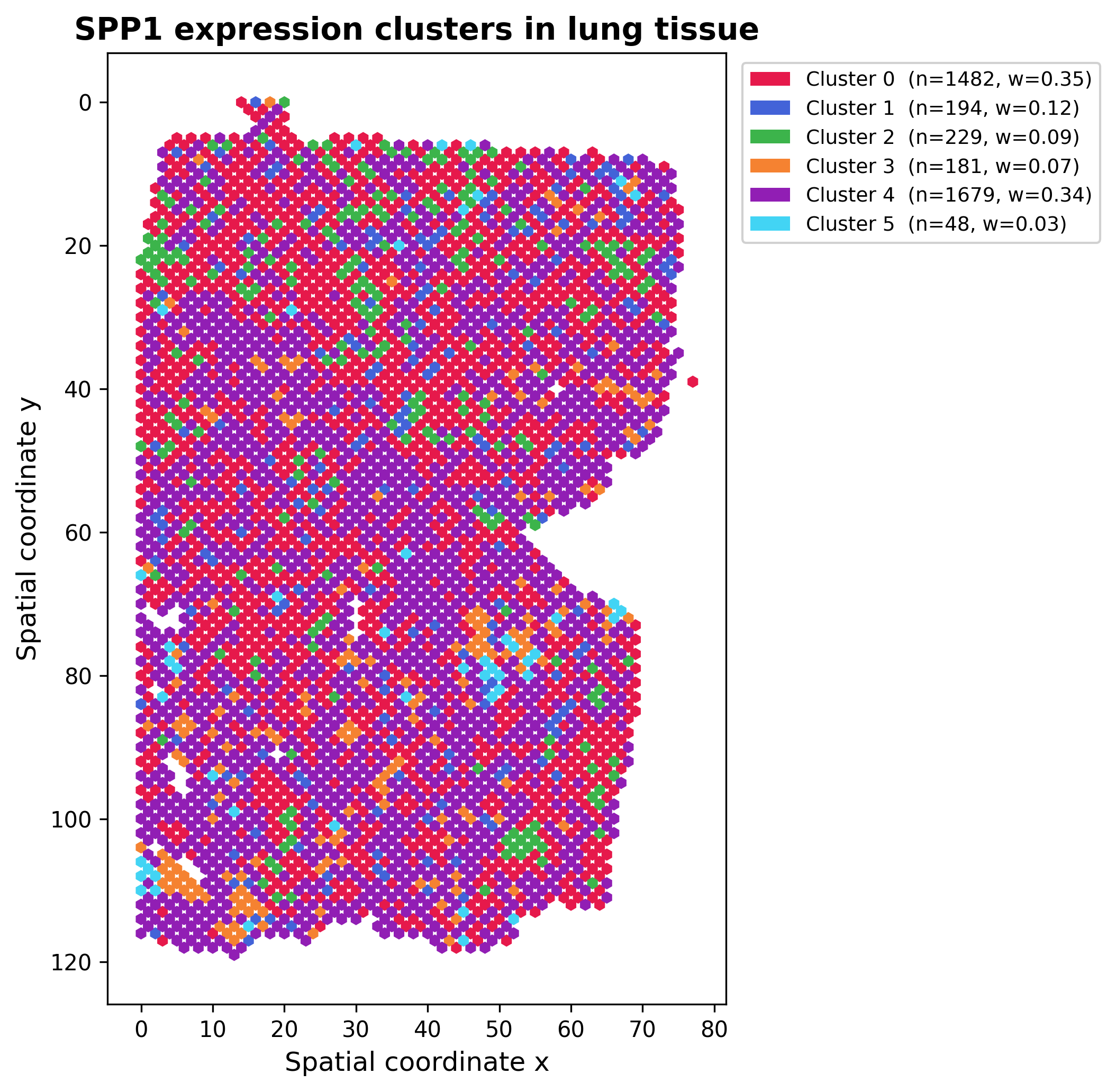}
    \caption{Estimated cluster assignments mapped onto the spatial coordinates of the lung adenocarcinoma tissue section. Each spot is colored by its Dahl-partition cluster; spatially contiguous patches of the smaller clusters indicate biologically and geographically structured heterogeneity in the composition--\emph{SPP1} relationship.}
    \label{fig:trans_spatial}
\end{figure}

\begin{figure}[htbp]
    \centering
    \includegraphics[width=0.85\textwidth]{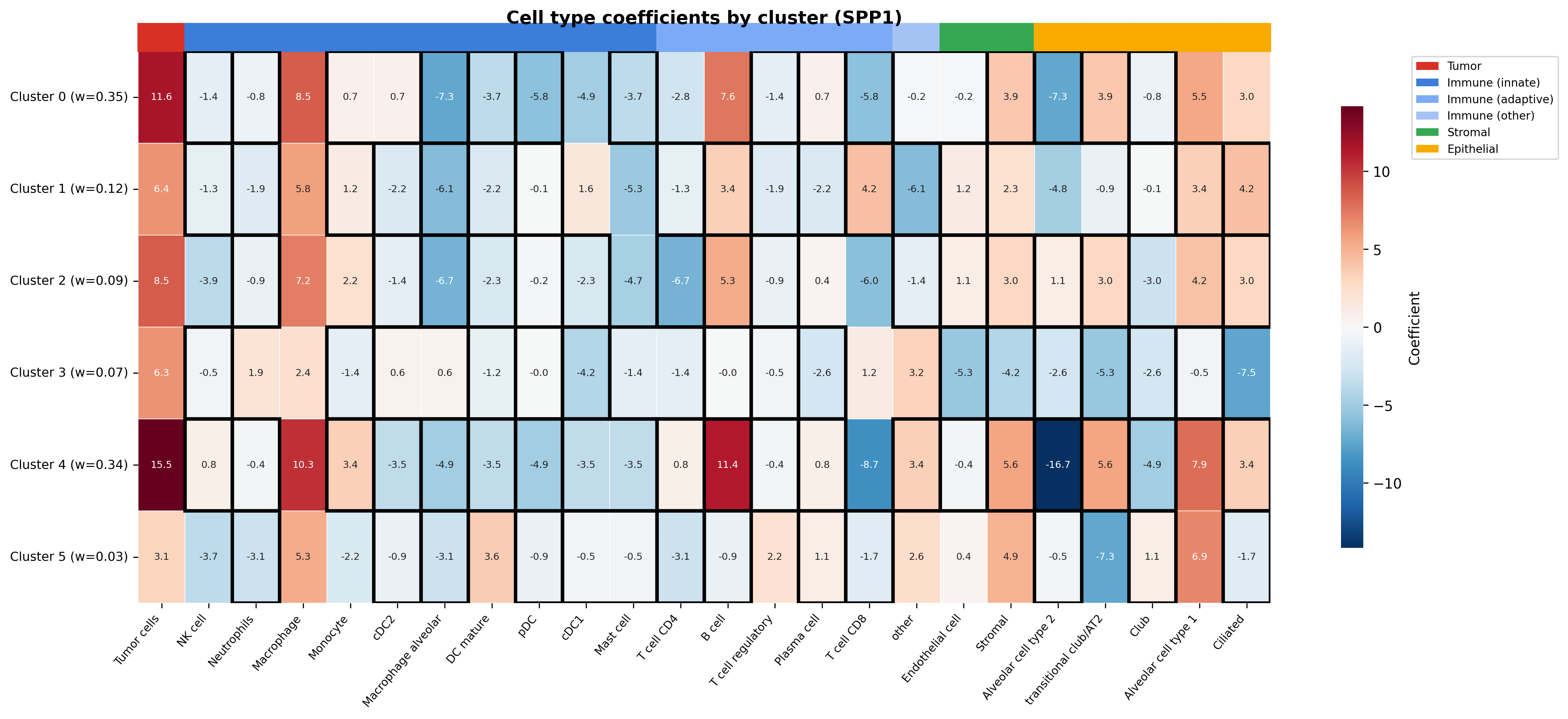}
    \caption{Fused cell type coefficients $\hat{\boldsymbol{\theta}}_k$ across the six \emph{SPP1} clusters. Cell types are grouped by lineage (tumor, immune--innate, immune--adaptive, immune--other, stromal, epithelial); black rectangles enclose contiguous runs of cell types fused to identical coefficient values within a cluster.}
    \label{fig:trans_theta}
\end{figure}

When the estimated cluster assignments are mapped back to the physical tissue coordinates in Figure~\ref{fig:trans_spatial}, the smaller clusters appear as dense and spatially contiguous patches rather than isolated scattered spots. This spatial organization supports the interpretation that the inferred clusters reflect biologically meaningful regional heterogeneity in the tumor microenvironment. In particular, the model identifies local regions where the association between cell type mixture and \emph{SPP1} expression differs from the dominant tissue-wide patterns.
Figure~\ref{fig:trans_theta} shows the fused cell type coefficients $\hat{\boldsymbol{\theta}}_k$ across the six clusters. Tumor cells and macrophages exhibit consistently large positive associations with \emph{SPP1} expression across clusters. This agrees with the biological role of \emph{SPP1} as a marker connected to tumor-associated signaling and macrophage-rich microenvironmental remodeling. The consistency of these positive coefficients suggests that the main \emph{SPP1} signal is driven by tumor and innate immune components across multiple tissue regions.

At the same time, the remaining cell type effects vary substantially across clusters, revealing localized differences in the composition--expression relationship. For example, Cluster 4 shows a uniquely strong positive coefficient for ciliated epithelial cells, suggesting a region-specific association between epithelial composition and \emph{SPP1} expression. Cluster 6 displays a distinct pattern involving alveolar type 1 cells, indicating that some spatial regions may follow specialized epithelial or alveolar-related expression regimes. These cluster-specific deviations demonstrate the value of allowing $\boldsymbol{\theta}_k$ to vary across latent groups rather than imposing a single cell type effect profile across the whole tissue section.

The GLBP prior also produces substantial dimensionality reduction. Although the model begins with 24 cell type coefficients in each cluster, many related or weakly distinguishable cell types are fused to common coefficient values. For example, dendritic cell and T-cell populations are merged within the larger clusters, indicating that the data do not support estimating separate effects for each highly granular immune subtype in those regions. This fusion is especially useful in spatial transcriptomics, where deconvolved cell type proportions are often correlated and difficult to interpret one at a time.

Overall, the spatial transcriptomics analysis demonstrates that the proposed model can recover both dominant and localized composition--expression regimes. It explains \emph{SPP1} expression well, identifies spatially coherent clusters, and preserves biologically interpretable signals from tumor cells and macrophages while shrinking redundant cell type effects together. The result is a more stable and interpretable representation of how cell type composition relates to \emph{SPP1} expression across heterogeneous tumor tissue.

\section{Discussion}\label{sec:discuss}

We proposed a Bayesian framework for regression with compositional predictors that jointly addresses two recurring features of modern applications: latent heterogeneity in the composition--response relationship and the presence of collapsed effect patterns among compositional parts. The model is built on the relative-shift parameterization of \citet{li2023s}, so that compositional effects are interpreted through identifiable reallocations of mass between components, rather than through non-identifiable coefficient levels under the simplex constraint. Within each latent mixture component, we place a projection-induced generalized $\ell_1$-ball prior (GLBP) \citep{xu2023bayesian} on the compositional effects to encourage equi-sparsity in the contrast space. This prior yields exact coefficient ties and therefore data-adaptive aggregation of parts with indistinguishable effects, while propagating uncertainty in both the latent partition and the induced collapsed structure.

The empirical studies highlight the practical role of collapsed structures. In the GDP analysis, the incremental $R^2$ attributable to the compositional sector shares is modest relative to the explanatory power of the non-compositional macroeconomic covariates, which capture much of the global mean variation. Nevertheless, introducing contrast-level aggregation substantially improves posterior stability of the latent partition, as measured by the mean Variation of Information between posterior draws and the Dahl point estimate (1.61 versus 2.04 for the baseline). This reduction in VI indicates that the GLBP prior on sectoral contrasts yields a more tightly concentrated posterior over partitions at the same level of resolution ($K=4$ active clusters under both models). The pattern is consistent with the interpretation that sectoral compositions are most informative for distinguishing structural regimes rather than for improving global mean prediction: by encouraging equi-sparsity within components, the prior suppresses redundant cluster-level variation and concentrates posterior mass on the informative contrast directions. In the spatial transcriptomics study, the model similarly supports stable inference in a regime characterized by high-dimensional compositions, potential subtype heterogeneity, and the practical need to summarize effects at an aggregated cell-type group level.

Several limitations suggest directions for refinement and extension. First, while the GLBP construction provides a generic mechanism for inducing equi-sparsity in the identifiable contrast space, the resulting aggregation can depend on the choice of contrast operator $A$. Beyond pairwise-difference contrasts, it would be useful to incorporate application-specific structure (e.g., sector hierarchies or biological lineage/taxonomy) through structured choices of $A$ that favor scientifically meaningful collapses. Second, the current specification focuses on main effects of compositional predictors; extending the relative-shift formulation to allow \emph{interactions}---either among compositional parts (e.g., synergistic reallocations) or between compositional and non-compositional covariates---could broaden the scope of interpretable effect statements. Third, for spatial transcriptomics and related settings, it is natural to couple mixture membership with \emph{spatial clustering processes} (e.g., Markov random field or Gaussian-process-based dependence) to encourage locally coherent partitions and improve robustness when subtype boundaries are spatially structured. Finally, further study of \emph{model misspecification} is warranted, including robustness to imperfect mixture assumptions, deviations from linearity in the link, and errors in estimated compositions; developing diagnostic tools and robustified likelihood or prior formulations would strengthen practical reliability. Future work along these lines would expand both the modeling flexibility and the inferential robustness of heterogeneous relative-shift regression, while preserving the reallocation-based interpretability that motivates the approach.

\bibliographystyle{agsm}
\bibliography{ref}

\newpage
\appendix

\end{document}